\documentclass[12pt]{article}
\usepackage{lineno}
\usepackage{subfig}
\usepackage{amsmath}
\usepackage{amsfonts}
\usepackage{multicol}
\usepackage{multirow}
\usepackage{advdate}
\usepackage{graphicx}

\newcommand\pubnumber{ATL-PHYS-PROC-2019-009}
\newcommand\pubdate{\today}

\def\institute{Graduate School of Science\\
Kobe University, 1-1, Rokkodai-cho, Nada-ku, Kobe, 657-8501, JAPAN}
\def\support{\footnote{Copyright 2019 CERN for the benefit of the ATLAS Collaboration.\\
                       \ \ \ \ \ \ Reproduction of this article or parts of it is allowed as specified in the CC-BY-4.0 license}}

\def\Title#1{\begin{center} {\Large #1} \end{center}}
\def\Author#1{\begin{center}{ \sc #1} \end{center}}
\def\Address#1{\begin{center}{ \it #1} \end{center}}

\newcommand\pubblock{\rightline{\begin{tabular}{l} \pubnumber\\
         \pubdate  \end{tabular}}}
\newenvironment{Abstract}{\begin{quotation}  }{\end{quotation}}
\newenvironment{Presented}{\begin{quotation} \begin{center} 
             PRESENTED AT\end{center}\bigskip 
      \begin{center}\begin{large}}{\end{large}\end{center} \end{quotation}}

\def\beq{\begin{equation}}
\def\eeq#1{\label{#1}\end{equation}}
\def\eeqn{\end{equation}}

\def\beqa{\begin{eqnarray}}
\def\eeqa#1{\label{#1}\end{eqnarray}}
\def\eeqan{\end{eqnarray}}

\let\bar=\overbar

\def\Dslash{\not{\hbox{\kern-4pt $D$}}}
\def\dslash{\not{\hbox{\kern-2pt $\del$}}}

\def\msb{{\bar{\ssstyle M \kern -1pt S}}}

\begin{document}
\frenchspacing
\begin{titlepage}
\pubblock

\vfill
\Title{Top-antitop charge asymmetry measurements\\ in the dilepton channel with the ATLAS detector}
\vfill
\Author{Shogo Kido, on behalf of the ATLAS Collaboration\support}
\Address{\institute}
\vfill
\begin{Abstract}
We report a measurement of the charge asymmetry $A_C$ in top quark pair production with the ATLAS experiment.
The measurement focuses on dilepton channels ($ee$, $e\mu$, $\mu\mu$).
The data are unfolded to parton level at full phase space using a fully Bayesian unfolding method.
Inclusive and differential measurements of the charge asymmetry are performed.
\end{Abstract}
\vfill
\begin{Presented}
$11^\mathrm{th}$ International Workshop on Top Quark Physics\\
Bad Neuenahr, Germany, September 16--21, 2018
\end{Presented}
\vfill
\end{titlepage}
\def\thefootnote{\fnsymbol{footnote}}
\setcounter{footnote}{0}

\section{Introduction}
The large mass of the top quark ($m_{t}\sim173$\,GeV) suggests that it may play a special role in the Standard Model (SM), as well as in several beyond Standard Model (BSM) theories. 
Top quark has a very short lifetime ($\tau_{t}\sim10^{-25}$\,s) and decays before hadronization. 
This allows an experimental test of the properties of a bare quark.
Therefore, precise measurements of the top quark properties are very interesting. The production of $t\bar{t}$ pairs at the Large Hadron Collider (LHC) in proton–proton collisions is symmetric under charge conjugation at leading order (LO) in quantum chromodynamics (QCD).
At next-to-leading order (NLO) in QCD, an asymmetry arises from interference between different Feynman diagrams~\cite{ACintro}.
The interference between the Born and one-loop diagram of the $q\bar{q} \to t\bar{t}$ leads to a positive asymmetry value.
The interference between initial-state radiation (ISR) and final-state radiation (FSR) diagrams leads to a negative asymmetry value.
The charge asymmetry, $A_C^{t\bar{t}}$, is defined as the following formula, by using $\Delta|y| = |y(t)| - |y(\bar{t})|$,

\begin{equation}
A_C^{t\bar{t}} = \frac{N(\Delta|y|>0)-N(\Delta|y|<0)}{N(\Delta|y|>0)+N(\Delta|y|<0)}
\end{equation}
here $y(t)$ and $y(\bar{t})$ are reconstructed rapidities of the top and antitop quark.
$N(\Delta|y|>0)$ and $N(\Delta|y|<0)$ represent the number of events with positive and negative $\Delta|y|$, respectively.
In dileptonic events, the $A_C^{t\bar{t}}$ can be measured by two charged leptons, two b-tagged jets and missing transverse energy (MET) from $t\bar{t} \to W^{+}bW^{-}\bar{b} \to l^{+}\nu_{l}{b}l^{-}\bar{\nu_{l}}\bar{b}$ process.
The $A_C^{t\bar{t}}$ is measured inclusively and differentially as a function of 
the invariant mass of ${t\bar{t}}$ system ($m_{t\bar{t}}$), 
transverse momentum of ${t\bar{t}}$ system ($p^{t\bar{t}}_{\text{T}}$) and velocity of ${t\bar{t}}$ system ($\beta^{t\bar{t}}_{z}$).
At high $m_{t\bar{t}}$, there are various BSM theories predicting an enhancement of the asymmetry.  
At low $p^{t\bar{t}}_{\text{T}}$, the asymmetry is dominated by the positive contribution from Born and one-loop amplitude interference.
At high $p^{t\bar{t}}_{\text{T}}$, the interference of ISR and FSR amplitudes causes a negative asymmetry.
Thus, the measurements as a function of $p^{t\bar{t}}_{\text{T}}$ probes different sources of asymmetry.
At high $\beta^{t\bar{t}}_{z}$, the fraction of the $t\bar{t}$ production via $q\bar{q}$ annihilation is larger and the asymmetry is enhanced in a model independent way.

\section{Analysis strategy}
In dileptonic $t\bar{t}$ events, exactly two oppositly charged isolated leptons and an invariant mass $m_{ll} > 15$\,GeV are required, together with at least two jets. 
Three different final states are considered in this analysis: events with two electrons in the final state ($ee$),
with one electron and one muon ($e\mu$), and with two muons ($\mu\mu$).
In the $e\mu$ channels, the background contamination is much smaller than in the same flavour channels and the background can be further suppressed by requiring the scalar sum of the $p_{\text{T}}$ of the two leading jets and leptons ($H_{\text{T}}$) to be larger than $130$\,GeV. 
In the same-flavor channels ($ee$ and $\mu\mu$),  
the largest contribution to the background comes from the associated production of $Z$\,bosons with heavy-flavor jets, 
therefore events with same sign leptons, of invariant mass satisfying the condition $|m_{ll}-m_{Z}| > 10$\,GeV,  are rejected.
Moreover, the huge background from low mass Drell–Yan (DY) production affects the measur$t\bar{t}$.
To suppress the contribution from the DY production,
the missing transverse energy must be greater than $30$\,GeV and at least one of the jets must be b-tagged.
The background arising from misidentified and nonprompt (NP) leptons is determined using both MC simulation and data.
The dominant sources of these NP leptons are semileptonic $b$-hadron decays, photons reconstructed as electrons, and electrons from photon conversions.
$W$+\,jets, single-top quark, and diboson productions are taken into account for the estimation of this background.
The top and antitop quark momenta are reconstructed by solving the system of equations that relate the particle momenta.
In dileptonic events, two neutrinos are produced and escape undetected. Thus, an underconstrained system is obtained.
To reconstruct the top and antitop quark momenta, a kinematic (KIN) method~\cite{KINmethod} is used.
In order to calculate the $\Delta|y|$ distribution at parton level in the whole phase space, fully Bayesian Unfolding (FBU)~\cite{FBU} is used.
For the treatment of systematic uncertainties in the Bayesian inference approach, the marginalization~\cite{FBU} is used.
The systematic uncertainties related to detector modeling uncertainties and the estimation of the backgrounds are included in the marginalization. 
The signal modeling uncertainties are estimated by using the migration matrix obtained with the nominal $t\bar{t}$ MC sample to unfold $\Delta|y|$ distributions predicted by different generators and different injected asymmetries.

\section{Results}
\begin{figure}[htbp]
\centering
\includegraphics[keepaspectratio,scale=0.42]{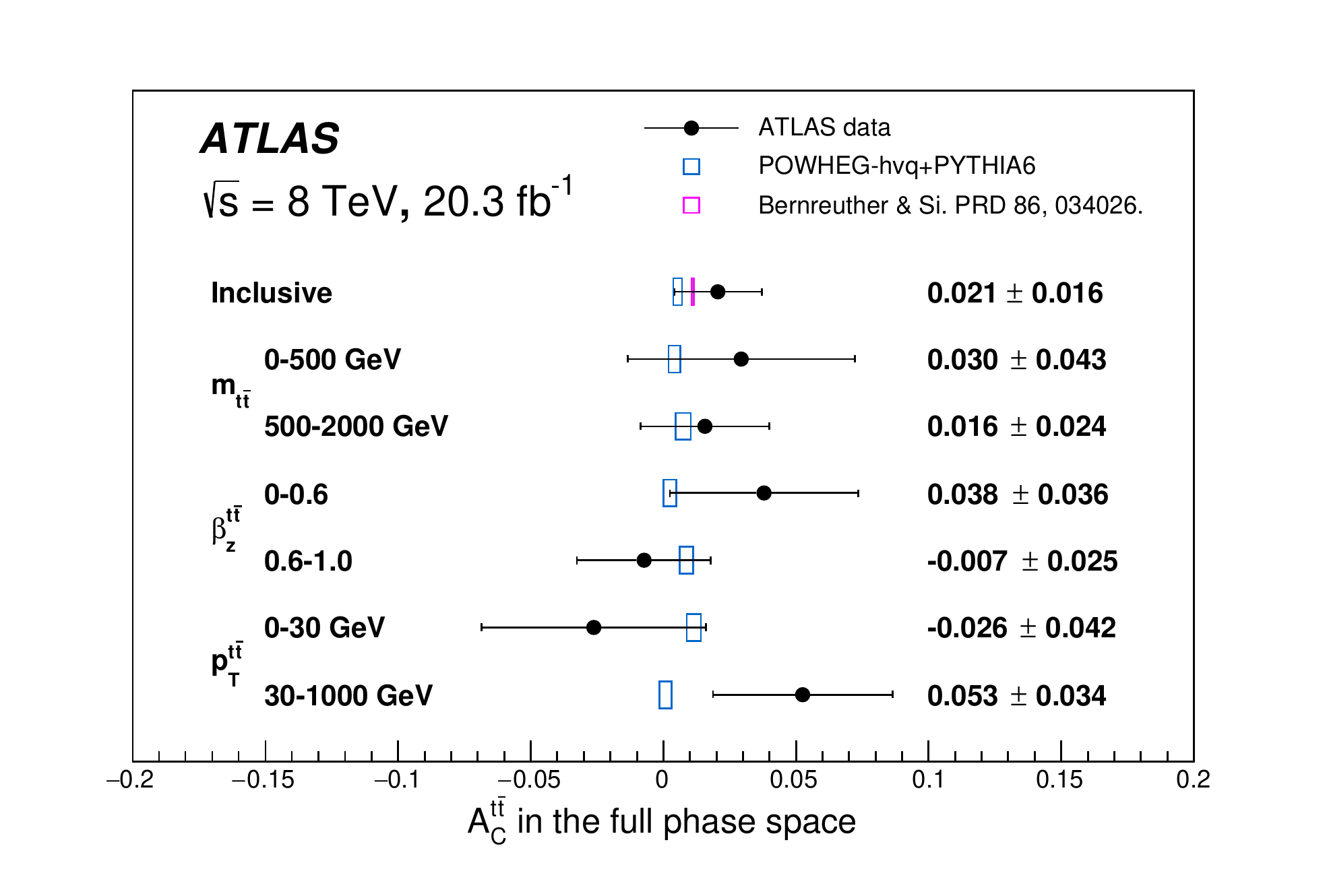}
\caption{Summary of all the measurements for the $t\bar{t}$ charge asymmetry in the full phase space~\cite{dileppub}. The predictions shown in blue are obtained using {\textsc Powheg-hvq}+{\textsc Pythia6} at NLO where the uncertainties are statistical, and the corresponding theoretical uncertainties are small compared to the experimental precision. The inclusive measurement in the full phase space is compared to a NLO+EW prediction~\cite{predictAc} shown in pink.}
\label{fig:summary_measure_Ac}
\end{figure}

The data sample used in this analysis corresponds to an integrated luminosity of 20.3\,fb$^{-1}$ at $\sqrt{s} = 8$\,TeV collected with the ATLAS detector~\cite{ATLASdetector}. 
The inclusive and differential results for the $t\bar{t}$ charge asymmetry in the full phase space is shown in Figure~\ref{fig:summary_measure_Ac}.
The measured inclusive value is $A_C^{t\bar{t}} = 0.021\pm0.016$.
The statistical uncertainty is the dominant contribution to the total uncertainty. 
The dominant systematic uncertainties across all the measurements are the signal modeling and the kinematic reconstruction uncertainty that is an intrinsic uncertainty of the reconstruction method due to the randomness in the smearing procedure.
The measurements at $\sqrt{s} = 8$ TeV are compatible with the SM prediction ($A_C^{t\bar{t}} = 0.0111\pm0.0004$~\cite{predictAc}) 
but cannot rule out the two typical BSM models (light octet, heavy octet)~\cite{BSMmodels}.
The measurements at $\sqrt{s} = 13$\,TeV are expected to significantly reduce both statistic and systematic uncertainties.

\section{Conclusion}
In this proceedings, the measurements of $t\bar{t}$ charge asymmetry at a centre-of-mass energy of $\sqrt{s} = 8$\,TeV was presented. 
Owing to the higher proton-proton center of mass energy, the data collected during the LHC run at $\sqrt{s} = 13$\,TeV are much more sensitive to BSM models, since they will allow to explore the region of high invariant mass of the $t\bar{t}$ system.
Therefore the ATLAS data set still to be analyses will open many interesting possibilities. 


\end{document}